\documentclass[journal,onecolumn]{ieeeconf}  
\IEEEoverridecommandlockouts                             
\usepackage{paralist}
\usepackage{amsfonts}
\usepackage{graphics} 
\usepackage{epsfig} 
\usepackage{mathptmx} 
\usepackage{times} 
\usepackage{amsmath} 
\usepackage{amssymb}  
\usepackage{physics}
\usepackage{mdwmath}
\usepackage{mdwtab}
\usepackage{color}
\usepackage[hidelinks]{hyperref}
\usepackage{algpseudocode}
\usepackage{algorithm}
\usepackage{caption}
\usepackage{subcaption}

\usepackage{xcolor}

\title{\LARGE \bf
Exploring Strange Entanglement: Experimental and Theoretical Perspectives on Neutral Kaon Systems}

\author{Nahid Binandeh Dehaghani, A. Pedro Aguiar, Rafal Wisniewski
\thanks{N. Dehaghani and A. Aguiar are with the Research Center for Systems and Technologies (SYSTEC), Electrical and Computer Engineering Department, FEUP - Faculty of Engineering, University of Porto, Rua Dr. Roberto Frias sn, i219, 4200-465 Porto, Portugal
        {\tt\small \{nahid,pedro.aguiar\}@fe.up.pt}}%
\thanks{R. Wisniewski is with Department of Electronic Systems, Aalborg University, Fredrik Bajers vej 7c, DK-9220 Aalborg, Denmark
        {\tt\small raf@es.aau.dk}}%
}

\begin{document}

\maketitle
\thispagestyle{empty}
\pagestyle{empty}
\begin{abstract} 
This chapter provides an in-depth analysis of the properties and phenomena associated with neutral K-mesons. Kaons are quantum systems illustrating strange behaviours. We begin by examining the significance of strangeness and charge parity violation in understanding these particles. The concept of strangeness oscillations is then introduced, explaining oscillations between $K^0$ and $\bar{K}^0$ states. The regeneration of $K_S$ is investigated, uncovering the underlying mechanisms involved. The discussion moves on to quasi-spin space, exploring its bases and their implications. The entangled states of kaon pairs $(K^0,\bar K^0)$ are considered, with a focus on maximally entangled neutral kaons and non-maximally entangled states. Decoherence effects on entangled kaons are examined, utilizing the density matrix description to capture the dynamics. A dedicated decoherence parameter is introduced to quantify the impact of decoherence. Furthermore, the chapter investigates the loss of entanglement through measures such as Von Neumann entanglement entropy, entanglement of formation, and concurrence. These measures provide insights into quantifying and characterizing entanglement in the context of neutral kaons. Through this comprehensive exploration of properties, phenomena, and entanglement dynamics, this chapter aims to pointing out recent works on neutral kaons, contributing to advancements in particle physics. 
\end{abstract}

\section{Introduction} 

Quantum entanglement, being among the most counterintuitive and subtle foundational elements of quantum mechanics, pertains to the correlations observed between distant components of certain composite systems. This intriguing phenomenon was brought to light by the pioneering work of Einstein, Podolsky and Rosen (EPR) in \cite{einstein1935can} and Schrödinger in \cite{schrodinger1935gegenwartige}, who uncovered a ``spooky'' characteristic of quantum machinery, better featured as nonlocality in the correlations of an EPR pair. Well-known and valuable tools for exploring this nonlocality are obtained by means of subsequent development of initial Bell inequalities, \cite{bell1987physics}, and their subsequently reformulated variations, \cite{clauser1974experimental, wigner1997hidden}. Experimental tests have consistently demonstrated the violation of Bell inequalities \cite{aspect1982experimental, tittel1998violation}, indicating the failure of local realistic theories and affirming the nonlocal nature of the universe.

Consequently, there is considerable interest in exploring the EPR-Bell correlations of measurements in various branches of physics, including particle physics. As a result, several pioneering researchers in particle physics have proposed investigating EPR-entangled massive particles, such as neutral kaons, \cite{feynman1963feynman, lee1981lee, okun2013leptons}. They referred to the unique characteristics of individual neutral kaon states, which exhibit various rare phenomena such as strangeness oscillation, small mass splitting, different lifetimes between the physical states, violations of two fundamental symmetries: charge parity (CP) and time reversal (T), regeneration when traversing a slab of material, and most notably, strange entanglement. Neutral kaons exhibit a unique form of entanglement known as strange entanglement, referring to the specific entanglement between two neutral kaons, \cite{bernabeu2022can}.

Numerous studies have been conducted to test quantum mechanics in the neutral kaon systems and search for CPT violation through neutral-meson oscillations. Notably, a significant focus on CP, T, and CPT violation in the neutral kaon system was first conducted at CERN in the CPLEAR experiment, \cite{apostolakis1998epr}. Additional contributions came from experiments such as NA48 and NA62, \cite{bizzeti2023recent}, which played key roles in discovering direct CP violation, yielding crucial experimental results. Moreover, experiments like KLOE, \cite{aloisio2002recent}, conducted at the DA$\Phi$NE collider, and its successor KLOE-2, \cite{Gauzzi:2019iks}, at the Frascati National Laboratory, achieved enhanced precision in investigating CPT violations and conducting quantum decoherence tests. The LHCb experiment, located at CERN's Large Hadron Collider (LHC), \cite{aaij2015lhcb}, and KOTO, performed at the Japan Proton Accelerator Research Complex (J-PARC), \cite{yamanaka2012the}, were amongst the other experiments which significantly contributed to our understanding of strange entanglement in neutral kaon system, providing valuable insights into the properties and behavior of entangled kaon states.

The investigation of the evolution of an entangled kaon system subjected to decoherence is a crucial aspect. This analysis is carried out by studying the system's behavior over time using the so-called master equation. As time progresses, the level of decoherence in the initially entangled kaon system increases, leading to a loss in the system's entanglement. This reduction in entanglement can be accurately assessed as in the field of quantum information, where the degree of entanglement in a state is quantified using specific measures, such as entropy of entanglement, concurrence, and entanglement of formation. These measures are widely employed for quantifying quantum entanglement. The unique characteristics of strange entanglement exhibited by neutral kaons, distinct from any other system, enable the exploration of novel phenomena. Considering the high importance of strange entanglement, a significant portion of the chapter is dedicated to exploring the stability of the entangled quantum system and examining the potential occurrence of decoherence due to interactions with its surrounding environment. We aim to understand the extent of these effects and their impact on entanglement by focusing on the correlation between decoherence and the loss of entanglement.

The structure of the chapter is as follows: In Section II, we review the strange behaviour of neutral kaons through several phenomena, i.e., strangeness and CP violation, strangeness oscillation, and regeneration. Section III introduces the bases in quasi–spin formalism, including the strangeness basis, free-space basis, and inside-matter basis. In Section IV, entangled states of kaon pairs are studied in two main subsections, i.e., maximally and non-maximally entangled states. In the next section, the effects of decoherence on entangled kaons are studied. Density matrix description of entangled kaon system is introduced and its evolution is studied through the Gorini-Kossakowski-Sudarshan-Lindblad equation. In section VI, the decline in entanglement is quantitatively measured, and explicit evidence of the loss of entanglement is provided. 
The chapter ends with conclusion and an overview on prospective research challenges.

\subsection{Notation} 
The ket symbol is used to denote a column vector in a Hilbert space, indicating the quantum state of the particle, i.e., $\left| {{K}^{0}} \right\rangle$ represents the state vector of a neutral kaon particle. The bra vector is used to describe the dual space or the bra state in quantum mechanics. In the case of a neutral kaon, $\langle {K^0} |$ represents the bra vector corresponding to the quantum state of the neutral kaon particle. We use the superscript $\dagger$ to show the conjugate transpose of a matrix (or vector). The symbol $\otimes$ indicates the tensor product, and $\oplus$ the direct sum. For $A$ and $B$ being operators, we use the notation $[A,B]=AB-BA$, and $\{A, B\} = AB + BA$. The expression ${A}(t={t_i}; {t_r})$ represents the state (or probability) $A$ at a specific time, where the time $t$ is equal to $t_i$, and $t_r$ represents a specific reference time. For a generic function $f(t)$, its derivative with respect to time is denoted by $\dot f (t)$. For a composite system consisting of subsystem $A$ and subsystem $B$, the partial trace over subsystem $B$ is denoted as $tr_B$. The imaginary unit is shown by $i = \sqrt{-1}$.

\section{Properties of the Neutral Kaons}
Kaons are the lightest strange mesons whose quark content is understood as $u\Bar{s}$, $s\Bar{u}$, $d\Bar{s}$ for the charged kaons $K^+(494 MeV/c^2)$ and $K^-(494 MeV/c^2)$, and the neutral kaon $K^0(498 MeV/c^2)$, respectively, \cite{martin2016particle}. Neutral K-mesons exhibit fascinating quantum phenomena that showcase their peculiar and intriguing behavior. In the following, we study notable instances that exemplify this strangeness.

\subsection{Strangness and Charge Parity Violation}

The neutral kaon and its antiparticle $\Bar{K}^0(498 MeV/c^2)=s\Bar{d}$ are distinguished by a quantum number $\mathcal{S}$, known as strangeness, such that $\mathcal{S}\left| {{K}^{0}} \right\rangle =+\left| {{K}^{0}} \right\rangle$, and $\mathcal{S}\left| {\Bar{K}^{0}} \right\rangle =-\left| {\Bar{K}^{0}} \right\rangle$. Kaons are pseudoscalar particles with a total spin of $0$ and parity $P=-1$ ($J^P=0^-$). They exhibit charge conjugation symmetry ($C$), which corresponds to the transformation ${{K}^{0}}\rightleftarrows {\Bar{K}^{0}}$.
Hence, for the joint transformation, one can write
\begin{equation*}
    CP\left| {{K}^{0}} \right\rangle =-\left| {{{\bar{K}}}^{0}} \right\rangle, \qquad CP\left| {{{\bar{K}}}^{0}} \right\rangle =-\left| {{K}^{0}} \right\rangle 
\end{equation*}
From a theoretical point of view, in order to obtain the CP eigenstates, one can implement the superposition of $\Bar{K}^{0}$ and ${{K}^{0}}$ as
\begin{equation*}
    \left| K_{1}^{0} \right\rangle =\frac{1}{\sqrt{2}}\left( \left| {{K}^{0}} \right\rangle -\left| {{{\bar{K}}}^{0}} \right\rangle  \right), \qquad \left| K_{2}^{0} \right\rangle =\frac{1}{\sqrt{2}}\left( \left| {{K}^{0}} \right\rangle +\left| {{{\bar{K}}}^{0}} \right\rangle  \right) \cdot
\end{equation*}
where $CP\left| K_{1}^{0} \right\rangle =+\left| K_{1}^{0} \right\rangle ,CP\left| K_{2}^{0} \right\rangle =-\left| K_{2}^{0} \right\rangle$. The state $K_{1}$ principally decays to two pions while $K_{2}$ primarily decays to three pions, which occurs around 600 times slower in comparison to the decay of $K_{1}$ into two pions. The reason why this decay occurs so slowly is due to the fact that the mass of $K_{2}$ is a bit greater than the total masses of the three pions. Strangness is not conserved in weak interactions, moreover, such interactions are CP violating. The observation of these two modes of decay led to the establishment of the existence of two weak eigenstates of the neutral kaons, called $K_L$ (K-long, $T$) and $K_S$ (K-short, $\theta$). The weak eigenstates are slightly different in mass, $\Delta m=m(K_L)-m(K_S)=3.49 \times 10^{-12} MeV$, however, they differ  considerably in their lifetimes and decay modes. The state $K_L$ ($K_S$) is a combination of $K_2$ ($K_1$) with a small portion of $K_1$ ($K_2$), expressed as
\begin{equation*}
   \left| {{K}_{S}} \right\rangle =\frac{1}{\sqrt{{{\left| p \right|}^{2}}+{{\left| q \right|}^{2}}}}\left( p\left| {{K}^{0}} \right\rangle -q\left| {{{\bar{K}}}^{0}} \right\rangle  \right) \qquad \qquad
  \left| {{K}_{L}} \right\rangle =\frac{1}{\sqrt{{{\left| p \right|}^{2}}+{{\left| q \right|}^{2}}}}\left( p\left| {{K}^{0}} \right\rangle +q\left| {{{\bar{K}}}^{0}} \right\rangle  \right) 
\end{equation*}
where $p=1+\epsilon$ and $q=1-\epsilon$, with $\epsilon$ being the complex CP violating parameter that is the same for $K_L$ and $K_S$, i.e., $\epsilon_L=\epsilon_S=\epsilon$. According to the CPT Theorem, \cite{greaves2014cpt}, since CP symmetry is violated, time reversal (T) symmetry must also be violated to maintain the overall CPT symmetry. The decay of $K_L$ (long-lived neutral kaon) is dominantly governed by CP violation, similar to the decay of $K_2$. The primary decay mode of $K_L$ is the three-pion decay, represented as $K_L\to 3\pi$, with a lifetime of approximately $\tau_L=5.17\times 10^{-8}$ seconds. Similarly, $K_S$ (short-lived neutral kaon) predominantly decays via the strong interaction, similar to the decay of $K_1$. The main decay mode of $K_S$ is the two-pion decay, denoted as $K_S\to 2\pi$, with a lifetime of approximately $\tau_S=8.954\times 10^{-11}$ seconds. It is important to note that while the dominant decay mode of $K_L$ is $K_L\to 3\pi$, there is also a small amount of CP-violating decay observed, specifically $K_L \to 2\pi$, \cite{griffiths2020introduction}.

\subsection{Strangeness oscillation}
\begin{figure}[t!]
  \centering
  \includegraphics[scale=0.62]{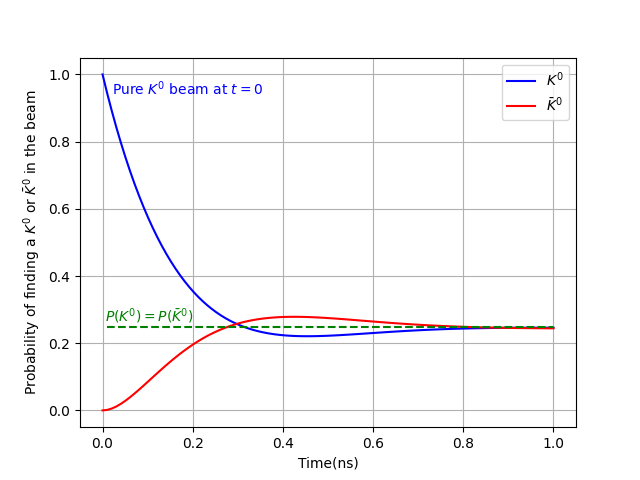}
  \caption{Probability of finding a $K^0$ or $\Bar{K}^0$ state in an initially produced $K^0$ beam over time }
  \label{probability}
\end{figure}

The two kaons $K^0$ and $\bar{K}^0$ transfer to common states, and subsequently they mix, meaning that they oscillate between $K^0$ and $\bar{K}^0$ before they decay. Consider the time dependent Schrödinger equation for the state vector $\left| {\psi }\left( t \right) \right\rangle$ as
\begin{equation*}
    i\hbar \left| \dot{\psi }\left( t \right) \right\rangle ={{H}}\left| \psi \left( t \right) \right\rangle
\end{equation*}
where $\hbar$ is the plank constant, and ${H}$ is the non-Hermitian effective mass Hamiltonian describing the decay characteristics and strangeness oscillations of kaons, defined as
\begin{equation}\label{Heff}
{H}=M-\frac{i}{2}\boldsymbol{\Gamma}  
\end{equation}
whose eigenstates are $K_S$ and $K_L$. The matrices $M$, related to mass, and $\boldsymbol{\Gamma}$, a decay-matrix, are $2\times 2$ Hermitian expressed as
\begin{equation*}
M=\left( \begin{matrix}
   {{M}_{11}} & {{M}_{12}}  \\
   M_{12}^{*} & {{M}_{11}}  \\
\end{matrix} \right), \qquad \boldsymbol{\Gamma}=\left( \begin{matrix}
   {{\Gamma}_{11}} & {{\Gamma}_{12}}  \\
   \Gamma_{12}^{*} & {{\Gamma}_{11}}  \\
\end{matrix} \right),
\end{equation*}
in which $\left\{\begin{aligned}
 & {{M}_{11}}=\frac{1}{2}\left( {{m}_{L}}+{{m}_{S}} \right) \\ 
 & {{M}_{12}}=\frac{1}{2}\left( {{m}_{L}}-{{m}_{S}} \right) \\ 
\end{aligned}\right.$ and $\left\{\begin{aligned}
  & {{\Gamma}_{11}}=\frac{\hbar}{2}\left( {{\Gamma}_{L}}+{{\Gamma}_{S}} \right) \\ 
 & {{\Gamma}_{12}}=\frac{\hbar}{2}\left( {{\Gamma}_{L}}-{{\Gamma}_{S}} \right) \\ 
\end{aligned}\right.$, with $\Gamma_{j}=\tau_j^{-1}$, $j=S,L$. The eigenvalues of ${H}$ satisfy 
\begin{equation*}
\left\{ \begin{aligned}
  & {{H}}\left| {{K}_{S}}\left( t \right) \right\rangle =\left( {{m}_{S}}-\frac{i\hbar }{2}{{\Gamma }_{S}} \right) \left| {{K}_{S}}\left( t \right) \right\rangle\\ 
 & {{H}}\left| {{K}_{L}}\left( t \right) \right\rangle =\left( {{m}_{L}}-\frac{i\hbar }{2}{{\Gamma }_{L}} \right) \left| {{K}_{L}}\left( t \right) \right\rangle\\ 
\end{aligned}\right.    
\end{equation*} The system evolves exponentially; i.e., the solutions to the Hamiltonian are obtained as
\begin{equation}\label{kskl}
    \left| {{K}_{S}}\left( t \right) \right\rangle ={{e}^{-(\frac{i}{\hbar }{{m}_{S}}+\frac{{{\Gamma }_{S}}}{2})t}}\left| {{K}_{S}}(t=0) \right\rangle, \qquad \left| {{K}_{L}}\left( t \right) \right\rangle ={{e}^{-(\frac{i}{\hbar }{{m}_{L}}+\frac{{{\Gamma }_{L}}}{2})t}}\left| {{K}_{L}}(t=0) \right\rangle 
\end{equation}
Subsequently, one can find the solution for ${K}^{0}$ and $\bar{K}^{0}$ as
\begin{equation*}
\begin{aligned}
  & \left| {{K}^{0}}\left( t \right) \right\rangle =\frac{1}{2}\left( {{e}^{-\left( \frac{i}{\hbar }{{m}_{S}}+\frac{{{\Gamma }_{S}}}{2} \right)t}}+{{e}^{-\left( \frac{i}{\hbar }{{m}_{L}}+\frac{{{\Gamma }_{L}}}{2} \right)t}} \right)\left| {{K}^{0}} \right\rangle +\frac{q}{2p}\left( -{{e}^{-\left( \frac{i}{\hbar }{{m}_{S}}+\frac{{{\Gamma }_{S}}}{2} \right)t}}+{{e}^{-\left( \frac{i}{\hbar }{{m}_{L}}+\frac{{{\Gamma }_{L}}}{2} \right)t}} \right)\left| {{{\bar{K}}}^{0}} \right\rangle  \\ 
 & \left| {{{\bar{K}}}^{0}}\left( t \right) \right\rangle =\frac{p}{2q}\left( -{{e}^{-\left( \frac{i}{\hbar }{{m}_{S}}+\frac{{{\Gamma }_{S}}}{2} \right)t}}+{{e}^{-\left( \frac{i}{\hbar }{{m}_{L}}+\frac{{{\Gamma }_{L}}}{2} \right)t}} \right)\left| {{K}^{0}} \right\rangle +\frac{1}{2}\left( {{e}^{-\left( \frac{i}{\hbar }{{m}_{S}}+\frac{{{\Gamma }_{S}}}{2} \right)t}}+{{e}^{-\left( \frac{i}{\hbar }{{m}_{L}}+\frac{{{\Gamma }_{L}}}{2} \right)t}} \right)\left| {{{\bar{K}}}^{0}} \right\rangle  \\ 
\end{aligned}
\end{equation*}
Suppose an experiment in which a beam of pure $K^0$ is produced at $t=t_i$, where $t_i$ is the initial time, via strong interaction. The probability of observing a $K^0$ in the beam at a subsequent time $t$ is determined by
\begin{equation*}
\begin{aligned}
 {{\left| \left\langle {{K}^{0}}\left| {{K}^{0}}\left( t \right) \right. \right\rangle  \right|}^{2}} & =\frac{1}{4}\left( {{e}^{^{\left( \frac{i}{\hbar }{{m}_{S}}-\frac{{{\Gamma }_{S}}}{2} \right)t}}}+{{e}^{^{\left( \frac{i}{\hbar }{{m}_{L}}-\frac{{{\Gamma }_{L}}}{2} \right)t}}} \right)\left( {{e}^{^{-\left( \frac{i}{\hbar }{{m}_{S}}+\frac{{{\Gamma }_{S}}}{2} \right)t}}}+{{e}^{^{-\left( \frac{i}{\hbar }{{m}_{L}}+\frac{{{\Gamma }_{L}}}{2} \right)t}}} \right) \\ 
 & =\frac{1}{4}\left( {{e}^{-{{\Gamma }_{S}}t}}+{{e}^{-{{\Gamma }_{L}}t}}+2{{e}^{-\frac{t}{2}\left( {{\Gamma }_{S}}+{{\Gamma }_{L}} \right)}}\cos \left( \frac{t\left( {{m}_{L}}-{{m}_{S}} \right)}{\hbar } \right) \right) \\ 
\end{aligned}
\end{equation*}
where the third term shows interference, which is the reason for an oscillation in the $K^0$ beam, \cite{martin2016particle}.
By following the same procedure, one can compute the probability of observing $\Bar{K}^0$ particles in a beam at a later time, given that the beam initially consists of ${K}^0$ particles. Therefore, 
\begin{equation*}
  {{\left| \left\langle {\Bar{K}^{0}}\left| {{K}^{0}}\left( t \right) \right. \right\rangle  \right|}^{2}}=\frac{1}{4}\frac{\left| q \right| ^2}{\left| p \right|^2}\left( {{e}^{-{{\Gamma }_{S}}t}}+{{e}^{-{{\Gamma }_{L}}t}}-2{{e}^{-\frac{t}{2}\left( {{\Gamma }_{S}}+{{\Gamma }_{L}} \right)}}\cos \left( \frac{t\left( {{m}_{L}}-{{m}_{S}} \right)}{\hbar } \right) \right) 
\end{equation*}
The $K^0$ beam oscillates with frequency $f=\frac{(m_L-m_S)}{2 \pi}$, with $(m_L-m_S) \tau_S=0.47$. The probability of finding a ${K}^0$ or $\Bar{K}^0$ from an initially pure ${K}^0$ beam is shown in \textbf{Figure \ref{probability}}. The plank constant $\hbar$ is considered as a unit for convenience. The oscillation becomes apparent when considering times on the order of a few $\tau_S$, before all $K_S$ mesons have decayed and only $K_L$ mesons remain in the beam. Therefore, in a beam initially consisting of only $K^0$ mesons at $t = 0$, the presence of the $\bar{K}^0$ is observed at a distance from the production source through its equal probability of being found in the $K_L$ meson. A similar phenomenon occurs when starting with a $\bar{K}^0$ beam, \cite{bertlmann2006entanglement}.
Over time, the composition of the beam undergoes variations in strangeness due to the different nature of ${K}^0$ and $\Bar{K}^0$ particles. This intriguing phenomenon is commonly referred to as strangeness oscillations, reflecting the oscillating strangeness content within the beam. In a broader context, this fascinating occurrence is known as flavor oscillations. 

\subsection{Regeneration of $K_S$}
A beam of K-meson decays in flight after a few centimeters, so the short-lived kaon state $K_S$ disappears, and only a pure beam of long-lived $K_L$ is left. By shooting the $K_L$ beam into a block of matter, which is usually regarded as a composition of protons
and neutrons for all practical purposes, then the ${K}^0$ and $\Bar{K}^0$ components of the beam interact dissimilarly with matter, which also causes the loss of quantum coherence between them. The $K^0$ particle engages in quasi-elastic scattering interactions with nucleons, while $\Bar{K}^0$ has the ability to produce hyperons. Since the emerging beam contains various different linear combinations of ${K}^0$ and $\Bar{K}^0$, i.e., a mixture of $K_L$ and $K_S$, the $K_S$ would eventually be regenerated in the beam, \cite{Pais1995note}.

\section{Bases in quasi–spin space}
The ``quasi-spin" picture for kaons, initially proposed by Lee and Wu, \cite{lee1966weak}, and later developed by Lipkin, \cite{lipkin1968cp}, offers notable advantages when compared to spin-$\frac{1}{2}$ particles or photons with vertical (V)/horizontal (H) polarization. The strangeness
eigenstates ${K}^0$ and $\Bar{K}^0$ are regarded as members of a quasi–spin doublet, where 
${K}^0=\left( \begin{matrix}
   1  \\
   0  \\
\end{matrix} \right)
$ (or V polarized photon) and $\Bar{K}^0=\left( \begin{matrix}
   0  \\
   1  \\
\end{matrix} \right)
$ (or H polarized photon) are considered as the quasi-spin states up ${{\left| \Uparrow  \right\rangle }_{z}}$ and down ${{\left| \Downarrow   \right\rangle }_{z}}$, respectively. 
All operators acting in the quasi–spin space can be expressed by Pauli matrices, i.e., ${\sigma}_x$, ${\sigma}_y$, ${\sigma}_z$. The strangeness operator $\mathcal{S}$ is identified by $\sigma_z$, i.e.,
\begin{equation*}
{\sigma}_z {\left| K^0  \right\rangle } = + {\left| K^0  \right\rangle }, \qquad {\sigma}_z {\left| \Bar{K}^0  \right\rangle } = - {\left| \Bar{K}^0  \right\rangle }, 
\end{equation*}
the CP operator with $-{\sigma}_x$, and the CP violation is relative to ${\sigma}_y$. This formalism is suitable for all two–level quantum systems. In this regard, the Hamiltonian in Eq. (\ref{Heff}) can be implemented as 
\begin{equation*}
H=\alpha I+\beta \left( \sin \theta {{\sigma }_{x}}+\cos \theta {\sigma}_y \right)
\end{equation*}
where $\alpha =\frac{1}{2}\left( {{m}_{L}}+{{m}_{S}}-\frac{i}{2}\left( {{\Gamma }_{L}}+{{\Gamma }_{S}} \right) \right)$, $\beta =\frac{1}{2}\left( {{m}_{L}}-{{m}_{S}}-\frac{i}{2}\left( {{\Gamma }_{L}}-{{\Gamma }_{S}} \right) \right)$, and the phase $\theta$ corresponds to the  CP parameter $\epsilon$ such that $e^{i\theta}=\frac{1-\epsilon}{1+\epsilon}$.

Overall, in the quasi–spin formalism, we may work with one of the following bases, \cite{bramon2007review}:
\begin{itemize}
    \item Strangeness basis $\{K^0,\Bar{K}^0\}$: 
    This basis is well-suited for examining electromagnetic and strong interaction processes that conserve strangeness, including the formation of $K^0\Bar{K}^0$ systems from non-strange initial states, for instance ${{e}^{+}}{{e}^{-}}\to \phi (1020)\to {{K}^{0}}{{\bar{K}}^{0}}$ or $p\bar{p}\to {{K}^{0}}{{\bar{K}}^{0}}$, and the detection of neutral kaons through strong kaon-nucleon interactions. This basis is orthonormal, i.e., $\left\langle  {{K}^{0}} | {{{\bar{K}}}^{0}} \right\rangle =0$.

    \item Free–space basis: $\{K_S,K_L\}$: 
    In the quasi–spin space, the weak interaction eigenstates are similar to the CP eigenstates $\left| {{K}_{1}} \right\rangle$ and $\left| {{K}_{2}} \right\rangle$. However, the $K_S$, $K_L$ basis provides a useful framework for analyzing the propagation of particles in free space while the CP basis is particularly suited for studying weak kaon decays. This basis is quasi–orthonormal with $\left\langle  {{K}_{S}} | {{K}_{S}} \right\rangle =\left\langle  {{K}_{L}} | {{K}_{L}} \right\rangle=1$, and $\left\langle  {{K}_{S}} | {{K}_{L}} \right\rangle =\left\langle  {{K}_{L}} | {{K}_{S}} \right\rangle=\frac{\epsilon+\epsilon^*}{1+{{\left| \epsilon  \right|}^{2}}}\simeq0$.

    \item Inside–matter basis: $\{K^\prime_S,K^\prime_L\}$: 
    The behavior of neutral kaons as they travel through a homogeneous medium of nucleonic matter, serving as both a regenerator and an absorber is determined by the medium Hamiltonian, which includes an extra strong interaction term as
    \begin{equation*}
    {{H}_{medium}}={{H}}-\frac{2\pi \nu }{{{m}_{K}}}\left( \begin{matrix}
   {{f}_{0}} & 0  \\
   0 & {{{\bar{f}}}_{0}}  \\
\end{matrix} \right)
    \end{equation*}
  where $\nu$ indicates the nucleonic density of the homogeneous medium, $m_K$ is the mean value of $K_{S,L}$ mass, $f_0$ and $\Bar{f}_0$ show the forward scattering amplitudes for $K^0$ and $\bar{K}^0$, respectively. The $\left| {{K^ \prime}_{L}} \right\rangle$ and $\left| {{K^ \prime}_{S}} \right\rangle$ are the eigenstates of ${H}_{medium}$ expressed as
  \begin{equation*}
\left| K_{L}^{'} \right\rangle =\frac{1}{\sqrt{1+{{\left| r\bar{\rho } \right|}^{2}}}}\left( \left| {{K}^{0}} \right\rangle +r\bar{\rho }\left| {{{\bar{K}}}^{0}} \right\rangle  \right), \quad 
 \left| K_{S}^{'} \right\rangle =\frac{1}{\sqrt{1+{{\left| r{{\left( {\bar{\rho }} \right)}^{-1}} \right|}^{2}}}}\left( \left| {{K}^{0}} \right\rangle -r{{\left( {\bar{\rho }} \right)}^{-1}}\left| {{{\bar{K}}}^{0}} \right\rangle  \right) 
 \end{equation*}
 where the dimensionless regenerator parameter $\rho$, the auxiliary parameter
 $\bar\rho$ and its inverse ${\left( {\bar{\rho }} \right)}^{-1}$ are introduced as
 \begin{equation*}
     \begin{aligned}
  & \rho \equiv \frac{\pi \nu }{{{m}_{K}}}\frac{{{f}_{0}}-{{{\bar{f}}}_{0}}}{{{m}_{L}}-{{m}_{S}}-\frac{i}{2}\left( {{\Gamma }_{L}}-{{\Gamma }_{S}} \right)} \\ 
 & \bar{\rho }\equiv \sqrt{1+4{{\rho }^{2}}}+2\rho, \quad {{\left( {\bar{\rho }} \right)}^{-1}}=\sqrt{1+4{{\rho }^{2}}}-2\rho  \\ 
\end{aligned}
 \end{equation*}
and $r=\frac{1-\epsilon}{1+\epsilon}$. This basis is also quasi–orthonormal. 
\begin{equation*}
\left\langle  K_{S}^{\prime} | K_{L}^{\prime} \right\rangle ={{\left\langle  K_{L}^{\prime} | K_{S}^{\prime} \right\rangle }^{*}}=\frac{1-{{\left| r \right|}^{2}}\left( {{{\bar{\rho }}}^{*}}/\bar{\rho } \right)}{\sqrt{1+{{\left| r\bar{\rho } \right|}^{2}}}\sqrt{1+{{\left| r/\bar{\rho } \right|}^{2}}}}
\end{equation*}

Two limiting cases exist:
  \begin{enumerate}
      \item For a very low density medium: $\left| {{K^ \prime}_{S}} \right\rangle \to \left| {{K}_{S}} \right\rangle$ and $\left| {{K^ \prime}_{L}} \right\rangle \to \left| {{K}_{L}} \right\rangle$
      \item For extremely high density media: $\left| {{K^ \prime}_{L}} \right\rangle \to \left| {\Bar{K}^{0}} \right\rangle$ and $\left| {{K^ \prime}_{S}} \right\rangle \to \left| {{K}^{0}} \right\rangle$
  \end{enumerate}
  
\end{itemize}

\section{Entangled states of kaon pairs}
In general terms, we classify a state as entangled when it cannot be expressed as a convex combination of product states, otherwise it is separable, \cite{bruss2002characterizing}.
Quantum entanglement, as a central feature of quantum mechanic, is a phenomenon where two or more particles can become correlated in such a way that the properties of one particle are immediately affected by the properties of the other particle, regardless of the distance between them. In quantum information, entanglement is regarded as a resource. Hence, one is interested in maximally entangled quantum states. To this end, we investigate the entangled states of kaon pairs in two main class, i.e., maximally and non-maximally entangled states.

\subsection{Maximally entangled neutral kaons}
The spin-singlet states, initially proposed by Bohm, are the most commonly studied and simplest form of bipartite states. These states involve a pair of spin-1/2 particles. In analogy to the standard Bohm state, we consider entangled states of $K^0\bar{K}^0$, \cite{bertlmann2006entanglement,bramon2004quantum}. In both cases of $\Phi-$resonance decays and s–wave proton–antiproton annihilation, the process begins at time $t= 0$ with an initial state denoted as $\left| \phi (0) \right\rangle$
with global spin, charge conjugation and parity ${{J}^{PC}}={{1}^{--}}$ expressed as
$ \left| \phi \left(t= 0 \right) \right\rangle =\frac{1}{\sqrt{2}}\left({{\left| {{K}^{0}} \right\rangle }_{l}}\otimes{{\left| {{{\bar{K}}}^{0}} \right\rangle }_{r}}-{{\left| {{{\bar{K}}}^{0}} \right\rangle }_{l}}\otimes{{\left| {{K}^{0}} \right\rangle }_{r}} \right)$, which can further be written in free–space basis as
\begin{equation}\label{ent}
    \left| \phi \left( t=0 \right) \right\rangle =\frac{1+{{\left| \varepsilon  \right|}^{2}}}{\sqrt{2}\left| 1-{{\varepsilon }^{2}} \right|}\left[ {{\left| {{K}_{S}} \right\rangle }_{l}}\otimes{{\left| {{K}_{L}} \right\rangle }_{r}}-{{\left| {{K}_{L}} \right\rangle }_{l}}\otimes{{\left| {{K}_{S}} \right\rangle }_{r}} \right]
\end{equation}
The neutral kaons separate and can be observed both to the left $(l)$ and right $(r)$ of the source. The weak interactions, which violate CP symmetry, come into play only in Eq. (\ref{ent}). It is worth noting that this state is both antisymmetric and maximally entangled in the two observable bases. Consequently, any measurements performed will consistently yield left–right anticorrelated outcomes. After production, the left-moving and right-moving kaons undergo evolution as described by Eq. (\ref{kskl}) for respective proper times $t_l$ and $t_r$. This formal evolution results in the formation of the ``two-times'' state.  Therefore,
\begin{equation}\label{entsl}
    \left| \phi \left( {{t}_{l}},{{t}_{r}} \right) \right\rangle =\frac{1}{\sqrt{2}}{{e}^{-\left( {{\Gamma }_{S}}{{t}_{l}}+{{\Gamma }_{L}}{{t}_{r}} \right)/2}}\left( {{\left| {{K}_{S}} \right\rangle }_{l}}\otimes{{\left| {{K}_{L}} \right\rangle }_{r}}-{{e}^{\left( i\Delta m+\Delta \Gamma /2 \right)\Delta t}}{{\left| {{K}_{L}} \right\rangle }_{l}}\otimes{{\left| {{K}_{S}} \right\rangle }_{r}} \right)
\end{equation}
where $\Delta t=t_l-t_r$, $\Delta m=m_L-m_S$, $\Delta \Gamma=\Gamma_L-\Gamma_S$, and $\epsilon \to 0$. Equivalently, Eq. (\ref{entsl}) can be written in strangeness basis 
\begin{equation*}
\begin{aligned}
    \left| \phi \left( {{t}_{l}},{{t}_{r}} \right) \right\rangle &=\frac{1}{2\sqrt{2}}{{e}^{-\left( {{\Gamma }_{S}}{{t}_{l}}+{{\Gamma }_{L}}{{t}_{r}} \right)/2}}\left( \left( 1-{{e}^{\left( i\Delta m+\Delta \Gamma /2 \right)\Delta t}} \right)\left( {{\left| {{K}^{0}} \right\rangle }_{l}}\otimes{{\left| {{K}^{0}} \right\rangle }_{r}}-{{\left| {{{\bar{K}}}^{0}} \right\rangle }_{l}}\otimes{{\left| {{{\bar{K}}}^{0}} \right\rangle }_{r}} \right) \right. \\    
   & +\left( 1-{{e}^{\left( i\Delta m+\Delta \Gamma /2 \right)\Delta t}} \right)\left( {{\left| {{K}^{0}} \right\rangle }_{l}}\otimes{{\left| {{{\bar{K}}}^{0}} \right\rangle }_{r}}-{{\left| {{{\bar{K}}}^{0}} \right\rangle }_{l}}\otimes{{\left| {{K}^{0}} \right\rangle }_{r}} \right)         
\end{aligned}
\end{equation*}
Typically, it is common to examine two-kaon states at a unique time, i.e., $t \equiv t_r = t_l$. In this scenario, we have the following equation
\begin{equation}\label{phit}
\begin{aligned}
   \left| \phi \left( t,t \right) \right\rangle &=\frac{1}{\sqrt{2}}{{e}^{-\left( {{\Gamma }_{S}}+{{\Gamma }_{L}} \right)t/2}}\left( {{\left| {{K}^{0}} \right\rangle }_{l}}\otimes{{\left| {{{\bar{K}}}^{0}} \right\rangle }_{r}}-{{\left| {{{\bar{K}}}^{0}} \right\rangle }_{l}}\otimes{{\left| {{K}^{0}} \right\rangle }_{r}} \right) \\ 
 & =\frac{1}{\sqrt{2}}{{e}^{-\left( {{\Gamma }_{S}}+{{\Gamma }_{L}} \right)t/2}}\left( {{\left| {{K}_{S}} \right\rangle }_{l}}\otimes{{\left| {{K}_{L}} \right\rangle }_{r}}-{{\left| {{K}_{L}} \right\rangle }_{l}}\otimes{{\left| {{K}_{S}} \right\rangle }_{r}} \right) \\ 
\end{aligned}
\end{equation}
exhibiting similar maximal entanglement and anti-correlations over time.

\subsection{Non–maximally entangled states}

In addition to the previously discussed maximally entangled state of kaons, there is interest in exploring other non-maximally entangled states for testing the local realism versus Quantum Mechanics theories. To prepare these states, we begin with the initial state described in Eq. (\ref{ent}). A thin and homogeneous regenerator is positioned along the right beam, as close as possible to the source of the two-kaon state. If the regenerator is placed in close proximity to this origin and the proper time ($\Delta t$) required for the right-moving neutral kaon to pass through the regenerator is sufficiently short, i.e. much smaller than $\tau_S$, weak decays can be neglected, and the resulting state after traversing the thin regenerator is obtained as
\begin{equation}\label{phidt}
\left| \phi \left( \Delta t \right) \right\rangle =\frac{1}{\sqrt{2}}\left( \left| {{K}_{S}} \right\rangle \otimes \left| {{K}_{L}} \right\rangle -\left| {{K}_{L}} \right\rangle \otimes \left| {{K}_{S}} \right\rangle +\eta \left( \left| {{K}_{S}} \right\rangle \otimes \left| {{K}_{S}} \right\rangle -\left| {{K}_{L}} \right\rangle \otimes \left| {{K}_{L}} \right\rangle  \right) \right)
\end{equation}
The regeneration effects is designated by $\eta =i\rho \left( \Delta m-\frac{i}{2} \Delta \Gamma \right)\Delta t$. One may note the difference of Eqs. (\ref{phit}) and (\ref{phidt}) at $t=0$ made by the terms linear in $\eta$. 
To intensify that difference, let the state Eq. (\ref{phidt}) propagate in free space up to a proper time $\tau_S \le T \le \tau_L$, so
\begin{equation}\label{phiT}
\begin{aligned}
    \left| \phi \left( T \right) \right\rangle = &\frac{{{e}^{-\left( {{\Gamma }_{L}}{{\tau }_{l}}+{{\Gamma }_{S}}{{\tau }_{r}} \right)/2}}}{\sqrt{2}}\left( \left| {{K}_{S}} \right\rangle \otimes \left| {{K}_{L}} \right\rangle -\left| {{K}_{L}} \right\rangle \otimes \left| {{K}_{S}} \right\rangle \right. \\
    & \left . -\eta \left( {{e}^{-\left( i\Delta m+\Delta \Gamma /2 \right)T}}\left| {{K}_{L}} \right\rangle \otimes \left| {{K}_{L}} \right\rangle -{{e}^{\left( i\Delta m+\Delta \Gamma /2 \right)T}}\left| {{K}_{S}} \right\rangle \otimes \left| {{K}_{S}} \right\rangle  \right) \right)       
\end{aligned}
\end{equation}
Equation (\ref{phiT}) shows that the $\left| {{K}_{L}} \right\rangle \otimes \left| {{K}_{L}} \right\rangle$ component has exhibited remarkable resilience against weak decays compared to the accompanying terms $\left| {{K}_{S}} \right\rangle \otimes \left| {{K}_{L}} \right\rangle$ and $\left| {{K}_{L}} \right\rangle \otimes \left| {{K}_{S}} \right\rangle$, resulting in its significant enhancement. Conversely, the $\left| {{K}_{S}} \right\rangle \otimes \left| {{K}_{S}} \right\rangle$ component has experienced substantial suppression and can therefore be disregarded provided that $T>>\tau_S$. By normalizing Eq. (\ref{phiT}) to the surviving pairs, one obtains
\begin{equation*}
  \left| \Phi  \right\rangle =\frac{1}{\sqrt{2+{{\left| {{R}_{L}} \right|}^{2}}+{{\left| {{R}_{S}} \right|}^{2}}}}\left( \left| {{K}_{S}} \right\rangle \otimes \left| {{K}_{L}} \right\rangle -\left| {{K}_{L}} \right\rangle \otimes \left| {{K}_{S}} \right\rangle +{{R}_{L}}\left| {{K}_{L}} \right\rangle \otimes \left| {{K}_{L}} \right\rangle +{{R}_{S}}\left| {{K}_{S}} \right\rangle \otimes \left| {{K}_{S}} \right\rangle  \right)  
\end{equation*}
in which ${{R}_{L}}=-r{{e}^{-\left( i\Delta m+\Delta \Gamma /2 \right)T}}$, ${{R}_{S}}=r{{e}^{\left( i\Delta m+\Delta \Gamma /2 \right)T}}$.
The state $\Phi$, which is non-maximally entangled, encompasses all pairs of kaons wherein both the left and right partners persist until the common proper time $T$. Due to the specific normalization of $\Phi$, kaon pairs exhibiting decay of one or both members prior to time $T$ need to be identified and excluded. This exclusion occurs before any measurement utilized in a Bell-type test, rendering this approach a ``pre-selection" procedure rather than a ``post-selection'' one, thereby avoiding any conflicts between local realism and quantum mechanics. 
Upon the establishment of the state  $\left| \Phi  \right\rangle$, it becomes essential to examine alternative joint measurements on each corresponding pair of kaons when conducting a Bell-type test.

\section{Decoherence effects on entangled kaons}
\begin{figure}[t!]
  \centering
  \includegraphics[width=0.8\textwidth,height=15pc]{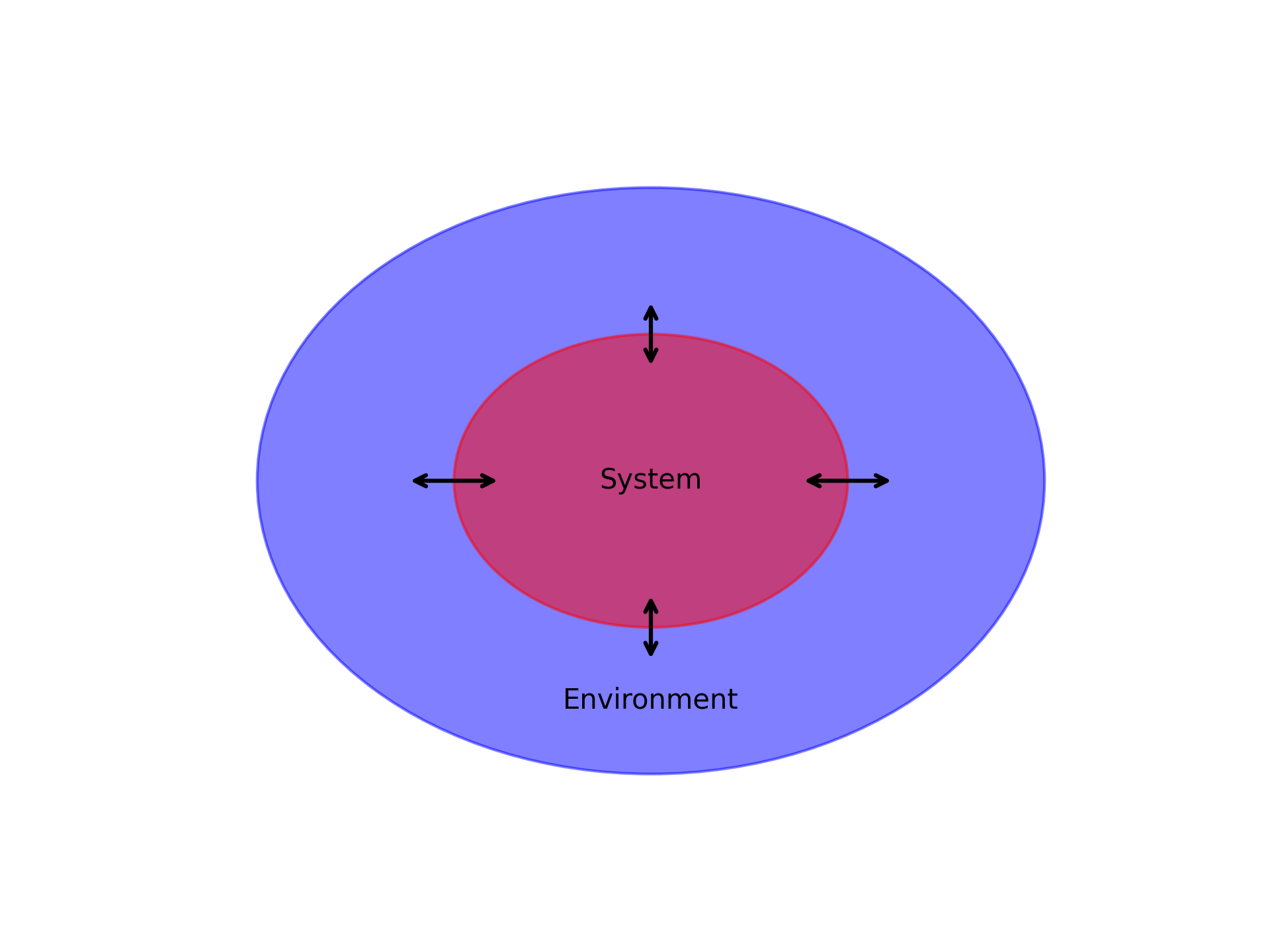}
  \caption{The overall system can be divided into two components: the system of interest, referred to as the ``system", and the surrounding ``environment". }
  \label{sysenv}
\end{figure}

Exploring the factors that could potentially lead to decoherence of entangled kaons is of high importance, \cite{bertlmann2006entanglement,di2009search,DiDomenico:2023kbu,bertlmann2003decoherence,syahbana2022study}. Besides, the decoherence allows us to gather insights into the quality of the entangled state. In the subsequent analysis, we explore potential decoherence effects that may arise from interactions between the quantum ``system" and its surrounding ``environment" as shown in \textbf{Figure \ref{sysenv}}.
Decoherence effects are mainly divided into two groups: standard and nonstandard, \cite{bertlmann2006entanglement}. Sources for ``standard" decoherence effects include:
\begin{itemize}
    \item Strong interaction scatterings of kaons with nucleons
    \item Weak interaction decays
    \item Noise from the experimental setup
\end{itemize}

Nonstandard decoherence effects are due to the fundamental modifications of quantum mechanics, such as:
\begin{itemize}
    \item Influence of quantum gravity \cite{rovelli2004quantum,kiefer2007quantum,gambini2007fundamental}
    \item Quantum fluctuations in the space-time structure at Planck mass scale \cite{sivaram2007special}
    \item Dynamical state-reduction theories \cite{pearle1989combining}
\end{itemize}

\subsection{Density matrix description of entangled kaon system}
We will now delve into the decoherence model within the Hilbert space $\mathcal{H}=\mathbb{C}^2$, which represents a two-dimensional complex vector space. Our analysis will specifically focus on the usual effective mass Hamiltonian, as denoted by Eq. (\ref{Heff}). Here, it is supposed that CP invariance is not violated as in the case of CPLEAR experiment, \cite{apostolakis1998epr}, whose data are not sensitive to the impacts of CP violation. Therefore, $p=q=1$ meaning that
\begin{equation*}
    \left| K_{1}^{0} \right\rangle \equiv \left| {{K}_{S}} \right\rangle , \quad \left| K_{2}^{0} \right\rangle \equiv \left| {{K}_{L}} \right\rangle , \quad \left\langle  {{K}_{S}} | {{K}_{L}} \right\rangle =0
\end{equation*}
We can effectively track the evolution of the density operator $\rho$ by employing the so-called Gorini-Kossakowski-Sudarshan-Lindblad equation representing the dynamics of a subsystem within a Markovian system as the entire system expressed as, \cite{dehaghani2022quantum},
\begin{equation}\label{Lind}
     \dot{\rho}(t)=\mathcal{L} \rho \left( t \right)
    =-i\left( H\rho \left( t \right)-\rho \left( t \right){{H}^{\dagger }} \right)+\underbrace{\sum\limits_{j}{\left( {{L}_{j}}\rho \left( t \right)L_{j}^{\dagger }-\frac{1}{2}\left\{ L_{j}^{\dagger }{{L}_{j}},\rho \left( t \right) \right\} \right)}}_{\mathcal{D}\left( \rho  \right)}
\end{equation}
in which $\mathcal{L}$ is the Liouville superoperator, and $L_j$ is the Lindblad (or jump) operator. The effect of decoherence is added by the dissipation term ${\mathcal{D}\left( \rho  \right)}$, for which we consider the following ansatz, \cite{bertlmann2003decoherence},
\begin{equation}\label{d}
    \mathcal{D}\left( \rho  \right)=\frac{\lambda }{2}\sum\limits_{j}{\left[ {{P}_{j}},\left[ {{P}_{j}},\rho  \right] \right]}\quad \text{with} \quad {{P}_{j}}=\left| {{K}_{j}} \right\rangle \left\langle  {{K}_{j}} \right|, \quad j=S,L
\end{equation}
where $\lambda\ge 0 $ is the decoherence parameter. Equation (\ref{d}) shows an especial case of dissipation where $L_j=\sqrt{\lambda} P_j$. Therefore, for the elements of density operator
$    \rho \left( t \right)=\sum\limits_{i,j=S,L}{{{\rho }_{ij}}\left( t \right)}\left| {{K}_{i}} \right\rangle \left\langle  {{K}_{j}} \right|$,
we attain 
\begin{equation*}
\begin{aligned}
  & {{\rho }_{SS}}(t)={{\rho }_{SS}}(0){{e}^{-{{\Gamma }_{S}}t}} \\ 
 & {{\rho }_{LL}}(t)={{\rho }_{LL}}(0){{e}^{-{{\Gamma }_{L}}t}} \\ 
 & {{\rho }_{LS}}(t)={{\rho }_{LS}}(0){{e}^{-\left( i({{m}_{L}}-{{m}_{S}})-\Gamma -\lambda  \right)t}} \\ 
\end{aligned}
\end{equation*}
where $\Gamma=\frac{1}{2}(\Gamma_S+\Gamma_L)$.

Consider the maximally entangled state Eq. (\ref{phit}) at initial time $t=0$ as
\begin{equation}\label{bell}
    \left| {{\psi }^{-}} \right\rangle =\frac{1}{\sqrt{2}}\left( \left| {{e}_{1}} \right\rangle -\left| {{e}_{2}} \right\rangle  \right)
\end{equation}
where $\left| {{e}_{1}} \right\rangle ={{\left| {{K}_{S}} \right\rangle }_{l}}\otimes {{\left| {{K}_{L}} \right\rangle }_{r}}$ and $\left| {{e}_{2}} \right\rangle ={{\left| {{K}_{L}} \right\rangle }_{l}}\otimes {{\left| {{K}_{S}} \right\rangle }_{r}}$. The total system Hamiltonian is then described by a tensor product of the one-particle Hilbert spaces as
   $ H=H_l \otimes I_r + I_l \otimes H_r $
with $l$ and $r$ denoting the direction of the moving particles. 
The state in Eq. (\ref{bell}) is a Bell state, \cite{nielsen2010quantum, kurgalin2021concise, zaman2018counterfactual}, and is equivalently expressed by the density operator
\begin{equation}\label{r0}
    \rho \left( 0 \right)=\frac{1}{2}\left( \left| {{e}_{1}} \right\rangle \left\langle  {{e}_{1}} \right|+\left| {{e}_{2}} \right\rangle \left\langle  {{e}_{2}} \right|-\left| {{e}_{1}} \right\rangle \left\langle  {{e}_{2}} \right|-\left| {{e}_{2}} \right\rangle \left\langle  {{e}_{1}} \right| \right)
\end{equation}
In this case, the projectors are $P_1=\left| {{e}_{1}} \right\rangle \left\langle  {{e}_{1}} \right|$ and $P_2=\left| {{e}_{2}} \right\rangle \left\langle  {{e}_{2}} \right|$, which project to the eigenstates of two-particle Hamiltonian. Therefore, the element wise time evolution obtained from Eq. (\ref{Lind}) with the ansatz Eq. (\ref{d}) is expressed as
\begin{equation*}
\begin{aligned}
  & {{{\dot{\rho }}}_{ij}}\left( t \right)=-2\Gamma {{\rho }_{ij}}\left( t \right) \qquad \quad \text{ for } i=j: {{\rho }_{ij}}\left( t \right)={{\rho }_{ij}}\left( 0 \right){{e}^{-2\Gamma t}} \\ 
 & {{{\dot{\rho }}}_{ij}}\left( t \right)=-\left( 2\Gamma +\lambda  \right){{\rho }_{ij}}\left( t \right) \quad \!\!\!\! \text{ for } i\ne j: {{\rho }_{ij}}\left( t \right)={{\rho }_{ij}}\left( 0 \right){{e}^{-\left( 2\Gamma +\lambda  \right)t}} \\ 
\end{aligned}
\end{equation*}
From Eq. (\ref{r0}), we already know $\rho_{11}(0)=\rho_{22}(0)=\frac{1}{2}$ and $\rho_{12}(0)=\rho_{21}(0)=-\frac{1}{2}$.
As a result, we acquire the time-varying density operator as the following:
\begin{equation}\label{rhot}
    \rho \left( t \right)=\frac{1}{2}{{e}^{-2\Gamma t}}\left( \left| {{e}_{1}} \right\rangle \left\langle  {{e}_{1}} \right|+\left| {{e}_{2}} \right\rangle \left\langle  {{e}_{2}} \right|-{{e}^{-\lambda t}}\left( \left| {{e}_{1}} \right\rangle \left\langle  {{e}_{2}} \right|+\left| {{e}_{2}} \right\rangle \left\langle  {{e}_{1}} \right| \right) \right)
\end{equation}
From Eq. (\ref{Lind}), it follows that $\rho(t)=e^{\mathcal{L}t}\rho(0)$. This means that the initial state $\rho(0)$ is transformed to $\rho(t)$ by the completely positive and trace preserving (CPTP) map $\mathcal{V}(t)=e^{\mathcal{L}t}$ generated by the superoperator $\mathcal{L}$, \cite{breuer2002theory,manzano2020short}.
Note that while $\mathcal{V}(t)$ should satisfy trace-preserving characteristics, the non-Hermitian nature of the system Hamiltonian results in a deviation from the property of trace preservation. Specifically, we observe that $tr(\rho(0))=1$ but $tr(\rho(t))=e^{-2 \Gamma t}$, in other words, $tr(\dot{\rho}(t))\ne 0$.
The issue arising from the non-Hermitian Hamiltonian in this particular system has been addressed in previous studies, \cite{caban2005unstable, bertlmann2006open}. By introducing certain modifications to the Hilbert space and the dynamical equation, one can effectively work with this Hamiltonian. Therefore, the Hilbert space $\mathcal{H}$ of the system is extended by adding the Hilbert space $\mathcal{H}_0$ which corresponds to the decay states resulting from the dissipation, so $\mathcal{H}_{tot}=\mathcal{H} \oplus \mathcal{H}_{0}$. 
By using the effective mass Hamiltonian defined in Eq. (\ref{Heff}), the dynamical equation is then expressed as
\begin{equation}\label{dyn}
    \dot{\rho }\left( t \right)=-i\left[ M,\rho \left( t \right) \right]-\left\{ \frac{1}{2}\boldsymbol{\Gamma} ,\rho \left( t \right) \right\}
\end{equation}
Let define $B: \mathcal{H} \to \mathcal{H}_0$, and $\boldsymbol{\Gamma}=B^\dagger B$, then we obtain $tr(\dot{\rho})=-tr(B^\dagger B \rho(t))\ne 0$. By adding $B \rho(t) B$ to Eq. (\ref{dyn}), such that $ \dot{\rho }\left( t \right)=-i\left[ M,\rho \left( t \right) \right]-\left\{ \frac{1}{2}\boldsymbol{\Gamma} ,\rho \left( t \right) \right\}+B \rho(t) B$, then $tr(\dot{\rho}(t))=0$, meaning that the trace of $\rho(t)$ is preserved.

\subsubsection{Purity}
Decoherence arises exclusively from the influence of the factor ${{e}^{-\lambda t}}$ on the off-diagonal elements. Hence, for $t=0$, the density operator corresponds to a pure state, however, for $t>0$ and $\lambda \ne 0$, the state $\rho(t)$ does not exihibit a pure state anymore. As time elapses and environmental factors come into play, the quantum effects, characterized by coherences, gradually fade away, giving rise to the phenomenon of decoherence.

In quantum mechanics and particularly in the field of quantum information theory, the purity of a quantum state described by the density operator is defined as $\mathcal{P}(t)=tr(\rho^2(t))$. The purity expresses a measure on quantum states, and provides information regarding the degree of mixture in a given state. Here, the purity of $\rho(t)$ is, \cite{syahbana2022study}, $$\mathcal{P}(t)=\frac{1}{2}{{e}^{-4\Gamma t}}\left( 1+{{e}^{-2\lambda t}} \right)$$
that is, for $t=0$, $\mathcal{P}(t)=1$, and for $t>0$, $\mathcal{P}(t)<1$. We can conclude that the evolution of $\rho(t)$ transfers a pure state to a mixed state, which is due to the occurrence of decoherence phenomenon. In addition, the purity of a normalized quantum state satisfies ${\displaystyle {\frac {1}{2}}\leq \mathcal{P}(t) \leq 1\,}$ for a state defined upon a 2 dimensional Hilbert space. However, in this case the value of $\mathcal{P}(t)$ can be less than ${\frac {1}{2}}$. It happens when one of the conditions of density operator is not satisfied for $t>0$, i.e., $tr(\rho(t))\ne 1$.

\subsection{Decoherence parameter associated to entangled kaon system}
\begin{figure}[t!]
  \centering
  \includegraphics[scale=0.4]{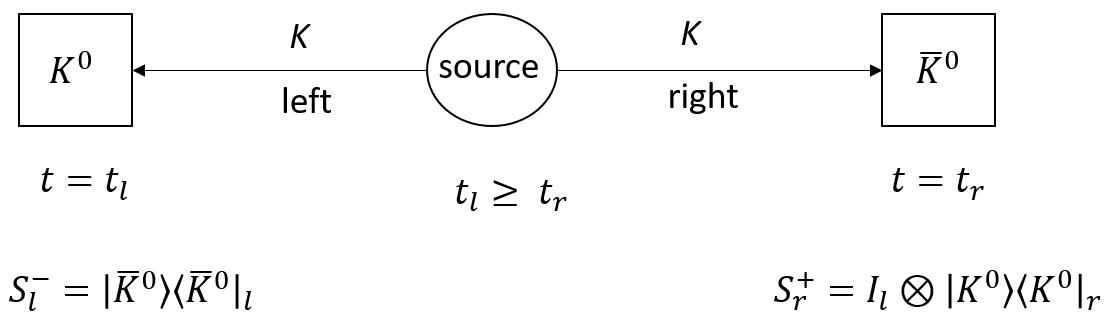}
  \caption{We consider a case that $\Bar{K}^0$ and ${K}^0$ are detected at the right side at $t=t_r$ and left side at $t=t_l$, respectively, where $t_l \ge t_r$. The operators $S^+_r$ and $S^-_l$ correspond to strangeness measurement at right and left sides.}
  \label{schematic}
\end{figure}

In the CPLEAR experiment, as described in \cite{apostolakis1998epr}, entangled kaons are generated. Subsequently, the strangeness content ($S$) of the right-moving and left-moving particles is measured at time $t=t_r$ and $t=t_l$, respectively. Consider a specific scenario where the detection reveals that a $\bar{K}^0$ is observed at the right side at time $t=t_r$, while a ${K}^0$ is detected at the left side at time $t=t_l$, where $t_r \le t_l$. We indicate two operators $S^+_r$ and $S^-_l$ to represent the measurement of strangeness at the right and left side, respectively (see \textbf{Figure \ref{schematic}}).

After the measurement is occurred at the right side, the density operator of the left moving particle turns out to be
\begin{equation}\label{rhol}
    {{\rho }_{l}}\left( t={{t}_{r}};{{t}_{r}} \right)=t{{r}_{r}}\left( S_{r}^{+}\rho \left( {{t}_{r}} \right) \right)
\end{equation}
From here, the probability of this case becomes, \cite{syahbana2022study},
\begin{equation}\label{expected}
    P\left( {{{\bar{K}}}^{0}},{{t}_{l}};{{K}^{0}},{{t}_{r}} \right)=tr\left( S_{l}^{-}{{\rho }_{l}}\left( {{t}_{l}};{{t}_{r}} \right) \right)=tr\left( t{{r}_{r}}\left( S_{r}^{+}{{\rho }_{l}}\left( {{t}_{r}} \right) \right) \right)
\end{equation}
In the following, we obtain $P\left( {{{\bar{K}}}^{0}},{{t}_{l}};{{K}^{0}},{{t}_{r}} \right)$. For the other cases, the same procedure can be employed. In order to determine Eq. (\ref{expected}), one can use $\left| {{K}_{S}} \right\rangle$ and $\left| {{K}_{L}} \right\rangle$ as the basis, and write Eq. (\ref{rhol}) as
\begin{equation}\label{r1}
 \!  \!  {{\rho }_{l}}\left( t={{t}_{r}};{{t}_{r}} \right)=\frac{1}{4}{{e}^{-2\Gamma {{t}_{r}}}}\left( \left| {{K}_{S}} \right\rangle {{\left\langle  {{K}_{S}} \right|}_{l}}+\left| {{K}_{L}} \right\rangle {{\left\langle  {{K}_{L}} \right|}_{l}}-{{e}^{-\lambda {{t}_{r}}}}\left( \left| {{K}_{S}} \right\rangle {{\left\langle  {{K}_{L}} \right|}_{l}}+\left| {{K}_{L}} \right\rangle {{\left\langle  {{K}_{S}} \right|}_{l}} \right) \right)
\end{equation}
Suppose the density operator related to a left moving particle is expressed as
\begin{equation}\label{r2}
 \begin{aligned}
    {{\rho }_{l}}\left( t;{{t}_{r}} \right)=&{{\rho }_{SS}}\left( t;{{t}_{r}} \right)\left| {{K}_{S}} \right\rangle {{\left\langle  {{K}_{S}} \right|}_{l}}+{{\rho }_{SL}}\left( t;{{t}_{r}} \right)\left| {{K}_{S}} \right\rangle {{\left\langle  {{K}_{L}} \right|}_{l}}\\
   & \qquad+{{\rho }_{LS}}\left( t;{{t}_{r}} \right)\left| {{K}_{L}} \right\rangle {{\left\langle  {{K}_{S}} \right|}_{l}}+{{\rho }_{LL}}\left( t;{{t}_{r}} \right)\left| {{K}_{L}} \right\rangle {{\left\langle  {{K}_{S}} \right|}_{l}}\\
 \end{aligned}
\end{equation}
which is supposed to evade decoherence for the time interval $t>t_r$. Therefore, it evolves according to
$$    {{\dot{\rho }}_{l}}\left( t;{{t}_{r}} \right)=-i\left( H{{\rho }_{l}}\left( t;{{t}_{r}} \right)-{{\rho }_{l}}\left( t;{{t}_{r}} \right){{H}^{\dagger }} \right)$$
Hence, the following equation is obtained
\begin{equation*}
 \begin{aligned}
  {{{\dot{\rho }}}_{l}}\left( t;{{t}_{r}} \right)& =-{{\Gamma }_{S}}{{\rho }_{SS}}\left( t;{{t}_{r}} \right)\left| {{K}_{S}} \right\rangle {{\left\langle  {{K}_{S}} \right|}_{l}}+\left( i\Delta m-\Gamma  \right){{\rho }_{SL}}\left( t;{{t}_{r}} \right)\left| {{K}_{S}} \right\rangle {{\left\langle  {{K}_{L}} \right|}_{l}} \\ 
 &\qquad -\left( i\Delta m+\Gamma  \right){{\rho }_{LS}}\left( t;{{t}_{r}} \right)\left| {{K}_{L}} \right\rangle {{\left\langle  {{K}_{S}} \right|}_{l}}-{{\Gamma }_{L}}{{\rho }_{LL}}\left( t;{{t}_{r}} \right)\left| {{K}_{L}} \right\rangle {{\left\langle  {{K}_{L}} \right|}_{l}} \\ 
\end{aligned}
\end{equation*}
Let assume that $C_{ij}$ with $i,j=S,L$ is constant, so
\begin{equation*}
\begin{aligned}
    &{{\rho }_{SS}}\left( t;{{t}_{r}} \right)={{C}_{SS}}{{e}^{-{{\Gamma }_{S}}t}},\qquad{{\rho }_{LL}}\left( t;{{t}_{r}} \right)={{C}_{LL}}{{e}^{-{{\Gamma }_{L}}t}},\\
    &{{\rho }_{SL}}\left( t;{{t}_{r}} \right)={{C}_{SL}}{{e}^{\left( i\Delta m-\Gamma  \right)t}},\qquad {{\rho }_{LS}}\left( t;{{t}_{r}} \right)={{C}_{LS}}{{e}^{-\left( i\Delta m+\Gamma  \right)t}}\\
\end{aligned}    
\end{equation*}
From our knowledge of Eqs. (\ref{r1}) and (\ref{r2}), the values of $C_{ij}$ and subsequently $\rho_{ij}(t;t_r)$ can be obtained. Finally, by replacing $t=t_l$, we attain $P\left( {{{\bar{K}}}^{0}},{{t}_{l}};{{K}^{0}},{{t}_{r}} \right)$.

Explicitly, by assuming that $\Delta t=t_l-t_r$, we have the following results
\begin{equation}\label{pk}
\begin{aligned}
   P\left( {{{{K}}}^{0}},{{t}_{l}};{\bar{K}^{0}},{{t}_{r}} \right)&=P\left( {{{\bar{K}}}^{0}},{{t}_{l}};{{K}^{0}},{{t}_{r}} \right)\\
   &=\frac{1}{8}{{e}^{-2\Gamma {{t}_{r}}}}\left( {{e}^{-{{\Gamma }_{S}}\Delta t}}+{{e}^{-{{\Gamma }_{L}}\Delta t}}+2{{e}^{-\lambda {{t}_{r}}}}\cos \left( \Delta m\Delta t \right){{e}^{-\Gamma \Delta t}} \right)   
\end{aligned}   
\end{equation}
\begin{equation}\label{pk1}
\begin{aligned}
   P\left( {{{{K}}}^{0}},{{t}_{l}};{{K}^{0}},{{t}_{r}} \right)&=P\left( {{{\bar{K}}}^{0}},{{t}_{l}};{\bar{K}^{0}},{{t}_{r}} \right)\\
   &=\frac{1}{8}{{e}^{-2\Gamma {{t}_{r}}}}\left( {{e}^{-{{\Gamma }_{S}}\Delta t}}+{{e}^{-{{\Gamma }_{L}}\Delta t}}-2{{e}^{-\lambda {{t}_{r}}}}\cos \left( \Delta m\Delta t \right){{e}^{-\Gamma \Delta t}} \right)   
\end{aligned}   
\end{equation}
Let consider $\Delta t =0$, i.e., $t_l=t_r=t$, then from Eq. (\ref{pk}), one obtains
\begin{equation*}
    P\left( {{{{K}}}^{0}},{{t}_{l}};{\bar{K}^{0}},{{t}_{r}} \right)=P\left( {{{\bar{K}}}^{0}},{{t}_{l}};{{K}^{0}},{{t}_{r}} \right)=\frac{1}{4}{{e}^{-2\Gamma {{t}}}}( 1-{e}^{-{\lambda}t})
\end{equation*}
which is in contradiction to the pure quantum mechanical EPR-correlations. The asymmetry of probabilities is the captivating factor of interest, as it directly responds to the interference term and can be quantified by means of experimental measurements. In the realm of pure quantum mechanics, where a system does not experience decoherence, we encounter this phenomenon by $A^{QM}$ expressed as
\begin{equation}\label{AQM}
\begin{aligned}
  {{A}^{QM}}\left( {{t}_{l}},{{t}_{r}} \right)& =\frac{P\left( {{K}^{0}},{{t}_{l}};{{{\bar{K}}}^{0}},{{t}_{r}} \right)+P\left( {{{\bar{K}}}^{0}},{{t}_{l}};{{K}^{0}},{{t}_{r}} \right)-P\left( {{K}^{0}},{{t}_{l}};{{K}^{0}},{{t}_{r}} \right)-P\left( {{{\bar{K}}}^{0}},{{t}_{l}};{{{\bar{K}}}^{0}},{{t}_{r}} \right)}{P\left( {{K}^{0}},{{t}_{l}};{{{\bar{K}}}^{0}},{{t}_{r}} \right)+P\left( {{{\bar{K}}}^{0}},{{t}_{l}};{{K}^{0}},{{t}_{r}} \right)+P\left( {{K}^{0}},{{t}_{l}};{{K}^{0}},{{t}_{r}} \right)+P\left( {{{\bar{K}}}^{0}},{{t}_{l}};{{{\bar{K}}}^{0}},{{t}_{r}} \right)} \\ 
 & =\frac{\cos \left( \Delta m\Delta t \right)}{\cosh \left( \frac{1}{2}\Delta \Gamma \Delta t \right)} \\ 
\end{aligned}
\end{equation}
in which $\Delta \Gamma = \Gamma_L - \Gamma_S$. In the decoherence model of entangled kaon system, since it is not known which particle will first be detected, $t_r$ in Eqs. (\ref{pk}) and (\ref{pk1}) needs to be replaced by $\tau=min(t_r,t_l)$. By inserting Eqs. (\ref{pk}) and (\ref{pk1}), we obtain
\begin{equation}\label{decmodel}
{{A}^{\lambda}}\left( {{t}_{l}},{{t}_{r}} \right)={{A}^{QM}}\left( {{t}_{l}},{{t}_{r}} \right) e^{-\lambda \tau} 
\end{equation}
which indicates that the decoherence effect represented by $e^{-\lambda \tau}$ is dependent on the time of the first detected kaon.

The decoherence model in Eq. (\ref{decmodel}) can be introduced in a phenomenological way, where the decoherence parameter $\lambda$ corresponds to an effective decoherence parameter $\zeta$ as $\zeta(t_l,t_r)=1-e^{-\lambda \tau}$.
Apparently, the value of $\zeta=0$ represents pure quantum mechanics, and $\zeta=1$ corresponds to complete decoherence or spontaneous factorization of the wave function (Schrödinger-Furry hypothesis). By means of standard least squares method, \cite{barlow1993statistics,bevan2013statistical}, $\zeta=0.13\pm 0.865$ is obtained in \cite{syahbana2022study}, which is in agreement with the results obtained from the effective variance method, where $\zeta=0.13_{-0.15}^{+0.16}$, \cite{bertlmann1999quantum,bertlmann2017quantum}, correspondent to $\lambda=(1.84_{+2.50}^{-2.17})\times 10^{-12}MeV$. The value of 
both parameters are compatible with quantum mechanics, i.e., $\zeta=0$ and $\lambda=0$ and far away from the total decoherence, i.e., $\zeta=1$ or $\lambda=\infty$. It indicates that the interaction between the system and its environment has negligible influence on the system. As a result, the quantum properties related to the entanglement of the strangeness are preserved without significant alteration.

\section{Loss of entanglement}
As time progresses, the degree of decoherence in the initially fully entangled $K^0 \bar{K}^0$ system increases, leading to a decrease in the system's entanglement. This loss of entanglement, defined as the disparity between an entanglement value and its maximum unity, can be precisely measured, \cite{bertlmann2006entanglement, hiesmayr2002puzzling}. In the realm of quantum information, the quantification of entanglement in a state is assessed by employing specific measures designed for quantifying entanglement. In this context, entropy plays a pivotal role. The entropy serves as a measure of the level of uncertainty or lack of knowledge associated with a quantum state. If the quantum state is pure, maximum information about the system is provided, however, mixed states only offer partial information. The entropy quantifies the extent to which maximal information is absent. In the following, we focus on the three most important and wildly-used entanglement measures: Von Neumann entanglement entropy, Entanglement of formation, and concurrence. When the interest is focused on the effect of decoherence, one needs to compensate for the decay up to time $t$ to attain a proper density operator for the kaon system. Therefore, we divide the state Eq. (\ref{rhot}) with its trace, i.e., $\rho_N(t)=\frac{\rho(t)}{tr(\rho(t))}$.

\subsection{Von Neumann entanglement entropy} 
Von Neumann's entropy of the quantum state $\rho_N(t)$ is expressed as
\begin{equation*}
   S(\rho_N(t))=-tr(\rho_N(t)log_{d} \rho_N(t))
\end{equation*}
where $d$ is the dimension of the Hilbert space, i.e., for the Hilbert space of a qubit $d=2$, hence $0<S(\rho_N(t))<1$. For $t=0$, the entropy is zero, meaning that the state is pure and also maximally entangled. However, as $t \to \infty$, the entropy approaches to one, i.e., the system become mixed. The Von Neumann entropy is a good criteria of entanglement, specifically for pure quantum states, \cite{bertlmann2006entanglement,bertlmann2003decoherence,varizi2017open}.

The bipartite von Neumann entanglement entropy can be defined as the von Neumann entropy of any of its reduced states, \cite{bennett1996mixed}. This definition holds true because both reduced states have the same value, as can be demonstrated through the Schmidt decomposition of the state with respect to the bipartition. Consequently, the outcome remains unchanged regardless of the specific reduced state chosen. Generally, two subsystems are maximally entangled when their reduced density operators are maximally mixed.  
In our case, the left- (subsystem $l$) and right- (subsystem $r$) propagating kaons are our subsystems. Therefore, the reduced density operators are defined as 
\begin{equation*}
  \rho _{N}^{l}\left( t \right)=t{{r}_{r}}\left( {{\rho }_{N}}\left( t \right) \right), \qquad
 \rho _{N}^{r}\left( t \right)=t{{r}_{l}}\left( {{\rho }_{N}}\left( t \right) \right) 
\end{equation*}
The von Neumann entropy of $\rho _{N}^{l}\left( t \right)$ (or $\rho _{N}^{r}\left( t \right)$) provides the uncertainty in the subsystem $l$ (or $r$) before measuring the subsystem $r$ (or $l$). In the case of kaon system, we have
\begin{equation*}
  S(  \rho _{N}^{l}\left( t \right))=S(  \rho _{N}^{r}\left( t \right))=1 \quad \forall t\ge 0
\end{equation*}
which are independent of decoherence parameter $\lambda$, meaning that the correlation stored in the entire system is lost to the environment and not to the subsystems. 

\subsection{Entanglement of formation}
Entanglement entropy is also known as the entanglement of formation for pure states.
It is possible to express any density matrix as a collection of pure states forming an ensemble ${{\rho }_{i}}=\left| {{\psi }_{i}} \right\rangle \left\langle  {{\psi }_{i}} \right|$, with each pure state having a corresponding probability $p_i$, that is $\rho=\sum\limits_{i}{{{p}_{i}}{{\rho }_{i}}}$. For mixed states, entanglement of formation can be generalized by defining a quantity minimized over all the ensemble realizations of the mixed state. For the kaon system, we have 
$${{E}_{f}}\left( \rho  \right)=\min \sum\limits_{i}{{{p}_{i}}S\left( \rho _{i}^{l} \right)}$$
The entanglement of formation can be simplified to $E_f(\rho)\ge \mathcal{E}\left( f\left( \rho  \right) \right)$ by introducing the lower bound of ${E}_f\left( \left( \rho  \right) \right)$, given by
$$\begin{aligned}
  & \mathcal{E}\left( f\left( \rho  \right) \right)=H\left( \frac{1}{2}+\sqrt{f\left( \rho  \right) \left( 1-f\left( \rho  \right)  \right)} \right)\quad\text{for} \quad f\left( \rho  \right) \ge \frac{1}{2} \\ 
 &\mathcal{E}\left(f\left(\rho\right)\right)=0\quad\quad\quad\quad\quad\quad\quad\quad\quad\quad\quad\quad \!\!\text{for} \quad f(\rho)<\frac{1}{2} \\ 
\end{aligned}
$$
where $f\left( \rho  \right)=\max \left\langle e\left| \rho  \right|e \right\rangle $,
known as the fully entangled fraction of $\rho$, is the maximum over all completely entangled states $\left| e \right\rangle$, and $$H(x)=-xlog_2x-(1-x)log_2(1-x).$$
For our model, the lower bound expressed for ${{E}_{f}}\left( \rho  \right)$ is saturated, i.e., $E_f(\rho)= \mathcal{E}\left( f\left( \rho  \right) \right)$. Therefore, one can calculate the entanglement of formation simply by computing $ \mathcal{E}\left( f\left( \rho  \right) \right)$, \cite{bertlmann2006entanglement}.

The fully entangled fraction of $\rho_N(t)$ is $ f(\rho_N(t))=\frac{1}{2}(1+e^{-\lambda t})\ge 0$. Therefore, the entanglement of formation for the $K^0\bar{K}^0$ is assessed in terms of $E_f(\lambda)$ as
\begin{equation*}
E_f(\lambda)=-\frac{1+\sqrt{1-{{e}^{-2\lambda t}}}}{2}{{\log }_{2}}\frac{1+\sqrt{1-{{e}^{-2\lambda t}}}}{2}-\frac{1-\sqrt{1-{{e}^{-2\lambda t}}}}{2}{{\log }_{2}}\frac{1-\sqrt{1-{{e}^{-2\lambda t}}}}{2}
\end{equation*}
From which one can obtain the entanglement loss $L_E(t)$ as 
\begin{equation}\label{lf}
    L_E(t)=1-E\left( {{\rho }_{N}}\left( t \right) \right)\simeq \frac{\lambda }{\ln 2}t=\frac{1}{\ln 2}\xi \left( t \right)=1.44\xi \left( t \right)
\end{equation}
approximated for small values of $\lambda$. Equation (\ref{lf}) shows that the entanglement loss equals the weighted amount of decoherence. 
\subsection{Concurrence}
Wootters and Hill, in their research publications \cite{wootters1998entanglement,hill1997entanglement,wootters2001entanglement}, discovered a relation between entanglement of formation and a measure known as concurrence. This connection allows the expression of entanglement of formation for a general mixed state $\rho$ of two qubits in terms of the concurrence as
\begin{equation*}
{{E}_{f}}\left( \rho  \right)=\mathcal{E}\left( C\left( \rho  \right) \right) =H\left( \frac{1}{2}+\frac{1}{2}\sqrt{1-{{C}^{2}}\left( \rho  \right)} \right) 
\end{equation*}
with $0\le C\left( \rho  \right)\le 1 $. As The function $\mathcal{E}\left( C\left( \rho  \right) \right)$ is monotonically increasing from $0$ to $1$ as $C(\rho)$ goes from $0$ to $1$.

\begin{figure}[t!]
  \centering
  \includegraphics[scale=0.35]{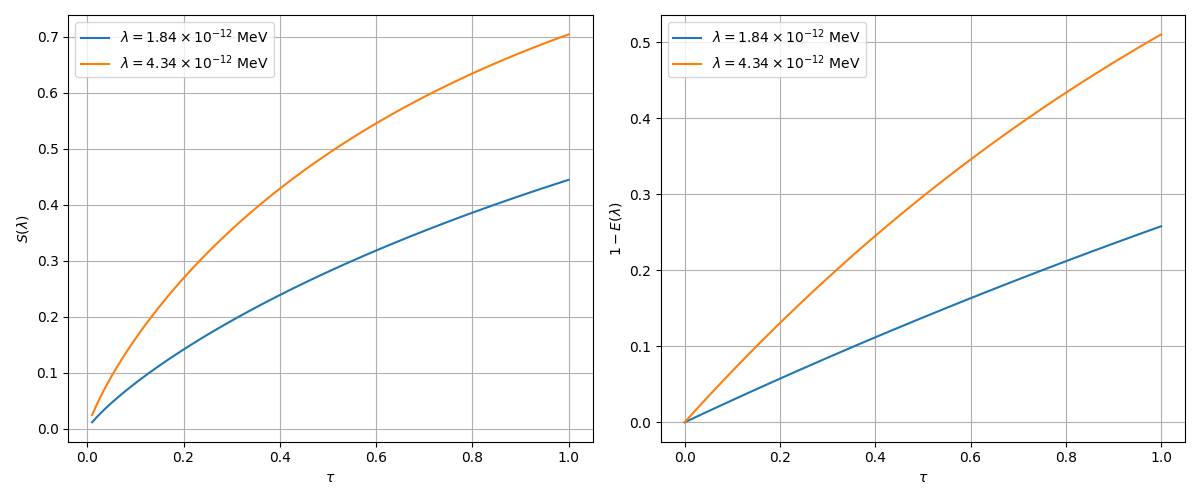}
  \caption{The time dependence of von Neumann entropy 
and the loss of entanglement of formation for two values of $\lambda$. The time $t$ is normalized versus $\tau_S$, i.e. 
$\tau=\frac{t}{\tau_S}$.}
  \label{loss}
\end{figure}

The concurrence $C(\rho)$ in defined as $C(\rho) \equiv max \{0,\lambda_1-\lambda_2-\lambda_3-\lambda_4\}$ with $\lambda_i$'s representing the square roots of the eigenvalues, in decreasing order, of $R=\rho \Tilde{\rho}$ matrix where $\Tilde{\rho}$ is the spin-flipped state of 
$\rho$ defined as
$${\tilde  {\rho }}=(\sigma _{{y}}\otimes \sigma _{{y}})\rho ^{{*}}(\sigma _{{y}}\otimes \sigma _{{y}})$$
The complex conjugate of $\rho$, i.e., $\rho ^{{*}}$ is taken in the basis $\left\{ \left| \Uparrow \Uparrow  \right\rangle ,\left| \Downarrow \Downarrow  \right\rangle ,\left| \Uparrow \Downarrow  \right\rangle ,\left| \Downarrow \Uparrow  \right\rangle  \right\}$.
In our model, since $\rho_N(t)$ is not variant under spin flip, hence $R=\rho^2_N$, and the concurrence is 
\begin{equation*}
    C(\rho_N(t)) \equiv max \{0,e^{-\lambda t}\}=e^{-\lambda t}
\end{equation*}
Hence, $L_C(t)$ is computed as
\begin{equation}\label{lc}
    L_C(t)=1-C\left( {{\rho }_{N}}\left( t \right) \right) = 1-e^{-\lambda t}=\xi (t)
\end{equation}
The value of $L_C(t)$ is precisely equivalent to decoherence $\xi (t)$, describing the factorization of the initial spin singlet state to the state ${{\left| {{K}_{S}} \right\rangle }_{l}}\otimes {{\left| {{K}_{L}} \right\rangle }_{r}}$ or ${{\left| {{K}_{S}} \right\rangle }_{l}}\otimes {{\left| {{K}_{L}} \right\rangle }_{r}}$. In both cases, Eqs. (\ref{lf}) and (\ref{lc}), show that the loss of entanglement is equivalent to decoherence, and increases linearly with time, \cite{bertlmann2003decoherence}. In \textbf{Figure \ref{loss}}, we show the loss of information by the von Neumann entropy $S(\lambda)$ in comparison with the loss of entanglement of formation $L_E(t)$ depending on the normalized time $\tau$ of the propagating $K^0\bar{K}^0$ system
for the experimental mean value $\lambda=1.84\times10^{-12}MeV$, and the upper bound $\lambda=4.34\times10^{-12}MeV$ of the decoherence parameter. The rate of increase in the loss of entanglement of formation is slower as time progresses, indicating the amount of resources required to create a specific entangled state. In the overall state, the level of entanglement decreases until separability is achieved, which occurs exponentially fast as time approaches infinity. In the CPLEAR experiment, where the propagation of one kaon for $2$ cm corresponds to the propagation time $\tau=0.55$, until an absorber measures it, the loss of entanglement is approximately $0.18$ and $0.38$ for the mean value and upper bound of $\lambda$, respectively. 

\section{Conclusions}
This chapter has provided a comprehensive analysis of the properties and phenomena associated with neutral K-mesons, highlighting their intriguing and often puzzling behaviors. We began by emphasizing the significance of strangeness and charge parity violation in the understanding of these particles. Next, the concept of strangeness oscillations, exemplified by the oscillations between $K^0$ and $\bar{K}^0$ states, was introduced and thoroughly explored. We delved into the regeneration of $K_S$ and unraveled the underlying mechanisms that govern these oscillations, shedding light on the intricate dynamics involved. Next, we examined the quasi-spin space and its bases, unraveling their implications and providing insights into the entangled states of kaon pairs, particularly focusing on both maximally and non-maximally entangled neutral kaons. This exploration has broadened our understanding of the entanglement properties exhibited by these particles. Furthermore, we dedicated significant attention to the study of decoherence effects on entangled kaons. Through the use of density matrix formalism, we captured the dynamic nature of decoherence and introduced a dedicated parameter to quantify its impact. Measures such as Von Neumann entanglement entropy, entanglement of formation, and concurrence were employed to explore the loss of entanglement, providing valuable tools for characterizing and quantifying entanglement in the context of neutral kaons. However, several intriguing questions for future research remain open. 

\subsection*{Outlook}
The study of kaon entanglement and decoherence raises a host of intriguing open questions that drive ongoing research and exploration. How do environmental interactions, such as scattering or absorption, impact the entanglement and coherence properties of neutral kaons? Can we develop methods to quantify and control these effects? Additionally, investigating entanglement dynamics in multipartite kaon systems, involving more than two neutral kaons, presents an exciting avenue of research. How does multipartite entanglement and its degradation relate to the underlying interactions and dynamics? Developing novel experimental techniques and theoretical models to explore these phenomena could shed light on the intricate nature of entanglement in kaon systems.

Furthermore, the generation and manipulation of entangled kaon states offer intriguing possibilities. Can we create specific entanglement patterns or preserve entanglement over extended timescales? Understanding and enhancing entanglement in kaon systems could have implications for quantum information processing and communication. The role of entanglement in comprehending and quantifying CP violation in kaon systems is another crucial aspect. How does entanglement contribute to our understanding of the underlying mechanisms driving CP violation? Exploring connections between the entanglement properties of neutral kaons and other areas of physics, such as quantum information theory, quantum field theory, or quantum gravity, holds promise for uncovering fundamental principles and phenomena.

Additionally, the presence of decoherence poses significant challenges. How does decoherence affect the entanglement properties of neutral kaons? Can we develop techniques to mitigate or minimize its detrimental effects and preserve entanglement over longer durations? Exploring the entanglement and coherence properties of neutral kaons within non-standard models beyond the Standard Model of particle physics could provide insights into new physics and phenomena. Moreover, extending the study of entanglement and decoherence to other meson systems or particles with similar characteristics is a compelling direction. How do the entanglement properties differ or align between different types of particles? Finally, experimental methodologies and theoretical frameworks for directly observing or measuring entanglement in kaon systems, as well as exploring the implications of entanglement for quantum computing and communication, open up exciting possibilities for practical applications in quantum technologies.

These open questions highlight the ongoing exploration and research endeavors, presenting intriguing avenues for further investigation and potential breakthroughs.
\section*{Acknowledgments}
The authors acknowledge the support of FCT for the grant 2021.07608.BD, the ARISE Associated Laboratory, Ref. LA/P/0112/2020, and the R$\&$D Unit SYSTEC-Base, Ref. UIDB/00147/2020, and Programmatic, Ref. UIDP/00147/2020 funds, and also the support of project, RELIABLE (PTDC/EEI-AUT/3522/2020) funded by national funds through FCT/MCTES. The work has been done in the honor and memory of Professor Fernando Lobo Pereira.

\end{document}